\documentclass[a4paper]{article}
\usepackage{latexsym}
\usepackage{color}
\begin {document}

\title {\bf From Fock's Transformation to de Sitter Space }
\author{T.~Foughali\footnote{Electronic address:
{\tt fougto\_74@yahoo.fr }}
\ and A.~Bouda\footnote{Electronic address:
{\tt bouda\_a@yahoo.fr}}\\
Laboratoire de Physique Th\'eorique, Facult\'{e} des Sciences Exactes, \\
Universit\'e de Bejaia, 06000, Bejaia, Algeria\\}

\date{\today}

\maketitle

\begin{abstract}
\noindent
As in Deformed Special Relativity, we showed recently that the Fock coordinate transformation can be
derived from a new deformed Poisson brackets. This approach allowed us to establish the corresponding
momentum transformation which keeps invariant the four dimensional contraction $p_{\mu} x^{\mu} $.
From the resulting deformed algebra, we construct in this paper the corresponding first Casimir.
After first quantization, we show by using the Klein-Gordon
equation that the spacetime of the Fock transformation is the de Sitter one. As we will see, the
invariant length representing the universe radius in the spacetime of Fock's transformation is exactly
the radius of the embedded hypersurface representing the de Sitter spacetime.
\end{abstract}

\vskip\baselineskip



\newpage

\section{Introduction}

Special relativity was modified in such a way to keep invariant, in addition to the speed of light, a
minimal length \cite{Amel-Pir, Ameliano1, Ameliano2, Mag-Smol1, Mag-Smol2, KMM} in the order of the Planck
length. The resulting theory is then called Deformed Special Relativity (DSR). The corresponding
coordinate transformation is not linear and it is proved later that it can be constructed by deforming
the Poisson brackets \cite{Ghosh-Pal}. By following the same way, we
showed recently \cite{Bou-Fou} that also the Fock coordinate transformation \cite{Fock},
\begin {equation}
t'={ \gamma \left( t-ux/c^{2} \right) \over \alpha_{R}}, \hskip6mm x'= {\gamma \left(x-ut\right) \over \alpha_{R}},
\hskip6mm y'={ y  \over \alpha_{R}}, \hskip6mm z'= {z  \over \alpha_{R}},
\end {equation}
where
\begin {equation}
\alpha_{R}= 1 + { 1 \over R } \left[ (\gamma-1)ct - \gamma { u x \over c} \right],
\end {equation}
$R$ being the universe radius and $\gamma=\left(1-u^{2}/c^{2} \right)^{-1/2}$, can be derived from a new
appropriate deformation of the Poisson brackets
\begin {eqnarray}
\{x^{\mu},x^{\nu}\} & = & 0 , \\
\{x^{\mu},p^{\nu}\} & = & -\eta^{\mu\nu}+{1 \over R}\eta^{0 \nu }x^{\mu}, \\
\{p^{\mu},p^{\nu}\} & = & -{1 \over R}\left(p^{\mu}\eta^{0\nu}-p^{\nu}\eta^{\mu 0}\right),
\end {eqnarray}
where $\eta^{\mu\nu} = (+1,-1,-1,-1)$. Here there are $c$ and $R$ which are invariant.
We stress that $c$ is a constant with a dimension of a velocity and it
represents the light speed only in the limit $R \rightarrow \infty$ \cite{Bou-Fou}.
The above brackets allowed us to establish the corresponding momentum transformation
\begin {equation}
E' = \alpha_{R} \gamma \left(E-up_{x} \right) , \ \ p'_{x} = \alpha_{R} \gamma \left(p_{x}-uE/c^{2}\right) ,
\ \ p'_{y} = \alpha_{R} p_{y} , \ \ p'_{z} = \alpha_{R} p_{z} ,
\end {equation}
which keeps invariant the four dimensional contraction $p_{\mu} x^{\mu} $. Contrary to earlier versions
\cite{KMM, Manida}, transformation (6) allows a coherent description of plane waves.
We observe that in the limit $R\rightarrow \infty$, (1) and (6) reduce to the Lorentz
transformations for the coordinates and the energy-momentum vector. We will call relations
(3), (4) and (5) "$R$-Minkowski phase space algebra".

Furthermore, we showed that the following expressions
\begin {equation}
I_{x} \equiv \left( 1 - { ct \over R} \right)^{-2} \eta_{\mu\nu}x^{\mu}x^{\nu}
\end {equation}
and
\begin {equation}
I_{p} \equiv \left( 1 - { ct \over R} \right)^{2} \eta_{\mu\nu}p^{\mu}p^{\nu}
\end {equation}
are invariant under transformations (1) and (6). We also observed that in
the limit $R\rightarrow\infty$, $I_{x}$ and $I_{p}$ reduce to the well-known invariants of
special relativity \cite{Bou-Fou}.

Our goal is to establish a link between the spacetime of Fock's coordinate transformation
and the one of de Sitter. First, we remark that $I_{p}$ is not a Casimir.
In fact, by using (4) and (5) we can check that it does not commute with the $R$-Lorentz
group generators $p^{0}$ and $p^{i}$. In order to construct the first Casimir of
the $R$-algebra, it is necessary to complete relations (3), (4) and (5) with others
between pure rotation,
\begin {equation}
M_{i}=\frac{1}{2}\epsilon_{ijk}J_{jk},
\end {equation}
and boost,
\begin {equation}
\tilde{N}_{i}=J_{0i},
\end {equation}
generators. In (9) and (10), $J_{\mu\nu} \equiv x_{\mu}p_{\nu}-x_{\nu}p_{\mu}$ represents the angular
momentum, ($\mu,\nu,...=0,1,2,3$, $i,j,...=1,2,3$) and $\epsilon_{ijk}$ is the Levi-Civita
antisymmetric tensor $(\epsilon_{123}=1)$. With the help of the above generators, it
is not yet possible to construct the first Casimir. In order to go further, we will follow
the method presented by Magpantay \cite{Mag1, Mag2}, where he developed the physics of the dual kappa
Poincar\'{e} algebra. Then, we will modify expression (10) for boost generator $\tilde{N}_{i}$ in such
a way to make a Casimir construction possible. That's what we will do in the next section. After first
quantization in section 3, we will show that the spacetime of the Fock transformation is identical to
the de Sitter one. Section 4 is devoted to conclusion.

\section{The Casimir construction}

We define new boost generators
\begin {equation}
N_{i} \equiv \tilde{N}_{i}-\frac{1}{2R}\eta_{\mu\nu}x^{\mu}x^{\nu}p_{i}
               = x_{0}p_{i} - x_{i}p_{0} - \frac{1}{2R}\eta_{\mu\nu}x^{\mu}x^{\nu}p_{i}
\end {equation}
which reproduce the usual ones in the limit $R\rightarrow \infty$.
We observe that the infinitesimal transformation of any function $O(x^{\mu},p^{\nu})$,
defined as in usual Lorentz transformation
\begin {equation}
\delta O= \{-{1\over2}\omega_{\mu\nu}J^{\mu\nu}, O\},
\end {equation}
is not affected by the additional term of $N_{i}$ with respect to $\tilde{N}_{i}$
because of the antisymmetric feature of the infinitesimal parameters $\omega_{\mu\nu}$.
Deformation (11) is reminiscent of the one proposed in \cite{Mag2}.
The $R$-algebra (3), (4) and (5) must be completed by the following brackets
\begin {eqnarray}
\{N_{i},p_{0}\} & = & -p_{i} + \frac{N_{i}}{R} , \\
\{N_{i},p_{j}\} & = & \eta_{ij} p_{0} - \frac{1}{R} \epsilon_{ijk}M_{k}, \\
\{M_{i},p_{0}\} & = & 0 , \\
\{M_{i},p_{j}\} & = & \epsilon_{ijk}p_{k}, \\
\{M_{i},M_{j}\} & = & \epsilon_{ijk}M_{k}, \\
\{M_{i},N_{j}\} & = & \epsilon_{ijk}N_{k}, \\
\{N_{i},N_{j}\} & = & - \epsilon_{ijk}M_{k},
\end {eqnarray}
which can be checked after a tedious calculation. We point out that relations (5) and (13)-(19)
constitute a particular case of the Bacry L\'{e}vy-Leblond algebras presented in \cite{Cac}.
As the Casimirs are scalars and observing that
\begin {equation}
M_{i}p_{i}= \epsilon_{ijk}x_{j}p_{k}p_{i} = 0,
\end {equation}
the first Casimir $C$ can be constructed by combining  $p_{\mu}p^{\mu}$, $M^{i}M^{i}$, $N^{i}N^{i}$,
$M^{i}N^{i}$ and $N^{i}p^{i}$
\begin {equation}
 C = p_{\mu}p^{\mu} + \alpha M^{i}M^{i} + \beta N^{i}N^{i} + \gamma N^{i}p^{i} + \lambda M^{i}N^{i},
\end {equation}
where the coefficients $\alpha$, $\beta$, $\gamma$ and $\lambda$ must be determined by imposing
to $C$ to commute with all the generators.

At this step, $x^{\mu}$ and $p^{\mu}$ will be substituted by the corresponding operators
$x^{\mu} \longmapsto \hat{x}^{\mu}$ and $p^{\mu} \longmapsto \hat{p}^{\mu} $.
Expression (11) of the generator $N_{i}$ is then ill defined because of the commutation relations
of $\hat{x}^{i}$ with $\hat{p}^{0}$ and $\hat{p}^{i}$. The symmetrization operation compel us to rewrite it as
\begin {equation}
N_{i}=\hat{x}_{0}\hat{p}_{i}-\frac{1}{2}\left( \hat{x}_{i}\hat{p}_{0} +
                              \hat{p}_{0}\hat{x}_{i} \right) -\frac{1}{2R} \hat{x}_{0}^{2}\hat{p}_{i} +
                                                    \frac{1}{4R}\left(\hat{x}^{j}\hat{x}^{j}\hat{p}_{i} +
                                                                       \hat{p}_{i}\hat{x}^{j}\hat{x}^{j}\right) .
\end {equation}
The Poisson brackets will be replaced by commutators in the following rule
\begin {equation}
 \{ \ \}\mapsto \frac{1}{i\hbar}[ \ ].
\end {equation}
The $R$-Minkowski phase space algebra now reads
\begin {eqnarray}
\left[\hat{x}^{\mu},\hat{x}^{\nu}\right] & = & 0 , \\
\left[\hat{x}^{0},\hat{p}^{0}\right] & = & -i\hbar \left(1- {\hat{x}^{0} \over R} \right), \\
\left[\hat{x}^{0},\hat{p}^{i}\right] & = & 0, \\
\left[\hat{x}^{i},\hat{p}^{0}\right] & = & i\hbar {\hat{x}^{i} \over R}, \\
\left[\hat{x}^{i},\hat{p}^{j}\right] & = & -i\hbar \eta^{ij}, \\
\left[\hat{p}^{i},\hat{p}^{j}\right] & = & 0, \\
\left[\hat{p}^{i},\hat{p}^{0}\right] & = & -i\hbar { \hat{p} ^{i}\over R},
\end {eqnarray}
and the resulting $R$-Poincar\'{e} algebra takes the form
\begin {eqnarray}
\left[N_{i},\hat{p}_{0}\right] & = & -i\hbar \hat{p}_{i} + i\hbar \frac{N_{i}}{R} , \\
\left[N_{i},\hat{p}_{j}\right] & = & i\hbar \ \eta_{ij} \hat{p}_{0} - \frac{i\hbar}{R} \epsilon_{ijk}M_{k}, \\
\left[M_{i},\hat{p}_{0}\right] & = & 0 , \\
\left[M_{i},\hat{p}_{j}\right] & = & i\hbar \epsilon_{ijk}\hat{p}_{k}, \\
\left[M_{i},M_{j}\right] & = & i\hbar \epsilon_{ijk}M_{k}, \\
\left[M_{i},N_{j}\right] & = & i\hbar \epsilon_{ijk}N_{k}, \\
\left[N_{i},N_{j}\right] & = & - i\hbar \epsilon_{ijk}M_{k},
\end {eqnarray}
Relations (31)-(37) obtained in the context of the Fock transformation are identical to that presented
in \cite{Mag1, Mag2} within the framework of the dual kappa Poincar\'{e} algebra.
Because of the above relations of commutations, expression (21) of the Casimir $C$ must be symmetrized.
Therefore, we write
\begin {equation}
 C = \hat{p}_{\mu}\hat{p}^{\mu} + \alpha M^{i}M^{i} + \beta N^{i}N^{i} +
            \frac{\gamma}{2} \left( N^{i}\hat{p}^{i} + \hat{p}^{i}N^{i} \right) +
                     \frac{\lambda }{2}\left( M^{i}N^{i} + N^{i}M^{i}\right).
\end {equation}
Imposing
\begin {equation}
 \left[C,\hat{p}_{0}\right] = 0
\end {equation}
leads to take $\beta = 0$, $\lambda = 0$ and $\gamma = 2/R$. Condition
\begin {equation}
 \left[C,\hat{p}^{i}\right] = 0
\end {equation}
gives $\alpha = -1/R^{2}$ and expression (38) turns out to be
\begin {equation}
 C = \hat{p}_{0}^{2} - \hat{p}^{i} \hat{p}^{i} - \frac{1}{R^{2}} M^{i}M^{i} + \frac{1}{R} \left( N^{i}\hat{p}^{i} + \hat{p}^{i}N^{i} \right) .
\end {equation}
One can check that
\begin {equation}
 \left[C,M^{i}\right] = 0,    \ \ \ \ \ \       \left[C,N^{i}\right] = 0 ,
\end {equation}
meaning that expression (41) of $C$ commute with all the generators of the $R$-Lorentz group. Of course,
in the limit $R \rightarrow \infty$, expression (41) reduces to the first Poincar\'{e} Casimir.
Contrary to DSR theory, we note that the Casimir depends on boost and rotation generators.
Identifying $R$ to $1/\kappa$, result (41) is identical to the one obtained by Magpantay \cite{Mag2}
in the context of the dual kappa Poincar\'{e} algebra.

\section{Towards the de Sitter spacetime}

Obviously, the usual representation
\begin {eqnarray}
\hat{x}^{\mu} & = & x^{\mu} , \\
\hat{p}^{\mu} & = &  i\hbar \frac{\partial}{\partial x_{\mu}} ,
\end {eqnarray}
does not work. We can check that the complete algebra (24)-(37) is satisfied if
we adopt the following representation:
\begin {eqnarray}
\hat{x}^{\mu} & = & x^{\mu} , \\
\hat{p}^{0} & = & i\hbar \left( \frac{\partial}{\partial x^{0}}- \frac{x^{\mu}}{R}  \frac{\partial}{\partial x^{\mu}}\right), \\
\hat{p}^{i} & = & - i\hbar \frac{\partial}{\partial x^{i}} .
\end {eqnarray}
We note that expression (46) differs from the one proposed by Magpantay \cite{Mag2}
by an additional term. From (46) and (47), we obtain
\begin {eqnarray}
\hat{p}^{0}\hat{p}^{0}= - \hbar^{2} \left[
             \frac{1}{R} \left( -1+\frac{x^{0}}{R} \right)\frac{\partial}{\partial x^{0}} +
             \frac{x^{i}}{R^{2}}\frac{\partial}{\partial x^{i}}
              + \left( 1 - \frac{x^{0}}{R} \right)^{2} \frac{\partial^{2}}{\partial \left( x^{0}\right)^{2}}
                                      \right.
            \hskip10mm&& \nonumber \\
                                      \left.
            - 2 \frac{x^{i}}{R} \left( 1 - \frac{x^{0}}{R} \right) \frac{\partial}{\partial x^{0}} \frac{\partial}{\partial x^{i}}
            + \frac{x^{i}x^{j}}{R^{2}} \frac{\partial}{\partial x^{i}} \frac{\partial}{\partial x^{j}}
                \right],
\end {eqnarray}
and
\begin {equation}
\hat{p}^{i} \hat{p}^{i} = - \hbar^{2}\triangle .
\end {equation}
Since
\begin {equation}
M^{i}  =  \frac{1}{2} \epsilon^{ijk} J^{jk} = \epsilon^{ijk} \hat{x}^{j}\hat{p}^{k},
\end {equation}
and taking into account relation
\begin {equation}
\epsilon^{ijk} \epsilon^{ils} = \delta^{jl}\delta^{ks} - \delta^{js} \delta^{kl},
\end {equation}
we obtain with the use of (28)
\begin {equation}
M^{i}M^{i}  =  - \hbar^{2} \left[
                   x^{i} x^{i} \triangle - x^{i}x^{j} \frac{\partial}{\partial x^{i}} \frac{\partial}{\partial x^{j}}
                   -2 x^{i} \frac{\partial}{\partial x^{i}}
                         \right].
\end {equation}
Relation (32) allows us to write
\begin {equation}
\hat{p}^{i}N^{i}  =  N^{i}\hat{p}^{i} + 3i\hbar \hat{p}^{0} .
\end {equation}
Using (22), we get to
\begin {eqnarray}
N^{i} \hat{p}^{i} + \hat{p}^{i} N^{i} = 2N^{i}\hat{p}^{i} + 3i\hbar \hat{p}^{0}  \hskip65mm&& \nonumber \\
                                       =  - \hbar^{2} \left\{
              2 \left[
              x^{0} - \frac{1}{2R}\left( \left(x^{0}\right)^{2} - x^{i}x^{i} \right)
                                                                                \right] \triangle
                                                                  \right.
            \hskip30mm&& \nonumber \\
                                                     \left.
            + 2 x^{i} \left[
             \left( 1 - \frac{x^{0}}{R}\right)\frac{\partial^{2}}{\partial x^{0}\partial x^{i}}
             - \frac{x^{j}}{R}\frac{\partial^{2}}{\partial x^{j}\partial x^{i}}
                                                                                                              \right] \right.
            \hskip10mm&& \nonumber \\
            \left.
             +3 \left[
             \left( 1 - \frac{x^{0}}{R}\right)\frac{\partial}{\partial x^{0}} - \frac{x^{i}}{R} \frac{\partial}{\partial x^{i}}
             \right]
                \right\}.
\end {eqnarray}
Substituting (48), (49), (52) and (54) in (41), expression of the first Casimir turns out to be
\begin {equation}
C = - \hbar^{2} \left( 1 - \frac{x^{0}}{R}\right)^{2}
       \left(\frac{\partial^{2}}{\left(\partial x^{0}\right)^{2}} - \triangle \right)
       - 2 \frac{\hbar^{2}}{R} \left( 1 - \frac{x^{0}}{R}\right) \frac{\partial}{\partial x^{0}} .
\end {equation}
It follows that the Klein-Gordon equation in $R$-spacetime, $C\phi = m^{2}c^{2}\phi$, takes the form
\begin {equation}
\left[ \hbar^{2} \left( 1 - \frac{x^{0}}{R}\right)^{2}
       \left(\frac{\partial^{2}}{\left(\partial x^{0}\right)^{2}} - \triangle \right)
       + 2 \frac{\hbar^{2}}{R} \left( 1 - \frac{x^{0}}{R}\right) \frac{\partial}{\partial x^{0}}
       + m^{2}c^{2} \right]\phi = 0 .
\end {equation}
We can check that this relation represents the Klein-Gordon equation in the de Sitter spacetime, known as
\begin {equation}
\left[ \frac{1}{\sqrt{|g|}} \partial_{\mu} \left( \sqrt{|g|} g^{\mu\nu} \partial_{\nu} \right)
                                                     + \frac{m^{2} c^{2}}{\hbar^{2}} \right]\phi = 0 ,
\end {equation}
with the following metric
\begin {equation}
ds^{2} = \frac{1}{\left( 1-x^{0}/R\right)^{2}} \left[ (dx^{o})^{2}-dx^{i}dx^{i}\right].
\end {equation}
This result means that the spacetime of the Fock transformation is the same to the de Sitter one
in its conformal metric. In fact, if we make the following coordinate transformation
\begin {equation}
x^{0} \rightarrow c\tau = x^{0} -R ,
\end {equation}
expression (58) takes the form
\begin {equation}
ds^{2} = \frac{R^{2}}{c^{2}\tau^{2}} \left[ c^{2}d\tau^{2}-dx^{i}dx^{i}\right],
\end {equation}
which is the conformal metric of the de Sitter spacetime \cite{Ib1,Ib2}. Furthermore, expression (60)
indicates that the invariant length $R$ representing the universe radius in the Fock transformation
spacetime is exactly the radius of the embedded hypersurface representing the de Sitter spacetime.
Also, we observe that if we compare (58) with the conformal metric presented in \cite{Ib2},
we deduce that the Hubble constant is equal to $H=c/R$. This observation strongly
reinforces the fact that $R$ is interpreted as the universe radius in the context of the Fock transformation.
These results are expected because of the presence of the parameter $R$ in the theory.
In fact, it is known that in vacuum the radius $R$ induces the presence of the cosmological constant $\Lambda$
since this latter is intimately linked to $R$ by $\Lambda = 3/R^{2}$. But precisely the solutions of Einstein's
equations in presence of the cosmological constant are the de Sitter space. This indicates that
the symmetry group of the model presented here is a de Sitter group \cite{AAMP} and the $R$-Minkowski spacetime
is indeed a de Sitter one.

We would like to add that other authors have already investigated
the de Sitter Special Relativity \cite{GHTWXZ,GHW,GHXZ}. There approach consists in covering the
de Sitter space by Beltrami coordinate patches which allow to express easly the law of inertia since
geodesics are represented by straight worldlines. Other approaches are also presented in \cite{Cac,AAMP}.

Also we mention that Magpantay \cite{Mag2} has shown that the spacetime of the dual DSR
is identical to the one of the de Sitter in planar coordinates. This result is different from ours
since our approach leads straightforwardly to the conformal metric of the de Sitter spacetime.
As indicated in \cite{Ib2}, it is the conformal metric in form (58) which is compatible with
astronomical observations.


\section{ Conclusion}

From the new deformed Poisson brackets which we recently proposed \cite{Bou-Fou} and used
to reproduce the Fock coordinate transformation, we constructed in this paper a complete set
of commutators of generators and the corresponding first Casimir.  Unlike in
DSR theory, this Casimir depends on boost and rotation generators. As in \cite{Mag2}, its
construction was made possible thanks to an appropriate redefinition of the boost generators.
After first quantization, we gave a realization of the corresponding deformed algebra and showed
by using the Klein-Gordon equation that the spacetime of the Fock transformation is identical to
the de Sitter one in its conformal metric, which is compatible with the astronomical observations \cite{Ib2}.
As we have seen, the invariant length representing the universe radius in the framework of Fock's
transformation is exactly the radius of the embedded hypersurface representing the de Sitter spacetime.
The same expression for the first Casimir and similar conclusions, with some nuances,
are obtained by Magpantay \cite{Mag2} within the framework of dual DSR. This means that the dual
kappa Poincar\'{e} algebra deals with the Fock transformation.
In the light of the results presented in this work, we strongly suggest that the
analogous of the Lorentz transformation in the Minkowski space is the Fock transformation in the
de Sitter space.

\bigskip
\bigskip

\noindent
{\bf REFERENCES}
\vskip\baselineskip

\end {document}